\documentclass[journal=jpccck,manuscript=article]{achemso}
\usepackage{tabularx}
\usepackage{booktabs}
\usepackage{setspace}
\author{Eduardo Cuervo Reyes}
\affiliation{Lab. of Inorg. Chem. ETH-Zurich, Switzerland}
\author{Elizabeth D. Stalder}
\affiliation{Lab. of Inorg. Chem. ETH-Zurich, Switzerland}
\author{Christian Mensing}
\affiliation{Lab. of Inorg. Chem. ETH-Zurich, Switzerland}
\author{Serhiy Budnyk}
\affiliation{Lab. of Inorg. Chem. ETH-Zurich, Switzerland}
\author{Reinhard Nesper}
\affiliation{Lab. of Inorg. Chem. ETH-Zurich, Switzerland}
\email{nesper@inorg.chem.ethz.ch}
\phone{+41 (0)44 632 30 69}
\fax{+41 (0)44 632 11 49}

\title[Unexpected Magnetism in {\AE}Si]
  {Unexpected Magnetism in Alkaline Earth Mono-Silicides}

\begin{document}
\begin{abstract}
Alkaline earth mono-silicides ({\AE}Si, {\AE} $=$ Ca, Sr, Ba) are  poor metals and their transport properties are not solely determined by the Zintl anion, in contrast to their Zintl-type composition. Their conducting network is  formed by the depopulated ${}^{1}_{\infty}$[Si$^{2-}$] $\pi$-system  and  {\AE}-$d$ states. This justifies the special local coordination of the metal atoms and the planarity of the silicon chains. The low density of carriers seems to be a playground for magnetic instabilities and the triangular prismatic arrangement of  {\AE} atoms responsible for the observed  weak glassy behavior.   
 
\end{abstract}

\section{Introduction}
Zintl phases (ZP) are main-group compounds which are located between  valence compounds and semi metallic elements according to their composition and properties. Classic ZP are formed by  alkaline (A) or alkaline-earth ({\AE}) metals\footnote{Some compounds  which also contain rare-earth elements are also considered to be Zintl phases.} together with one or more of the $p$-elements  which are metallic, semi metallic or  semiconducting (Ga, In, Tl, Si, Ge, Sn, Pb, As, Sb, Bi). These materials can be understood in terms of a formal ionic model describing the overall chemical bonding patterns\cite{Zintl1,Zintl2,Zintl3,Zintl4,Klemm,Mooser2,Klemm0,Busmann,Klemm2,Nesper2,Kautzlarich,Nesper0}. The most electropositive metal atoms form closed shell cations, as a consequence of  complete transfer of the valence electrons to lower lying states centered on the electronegative atoms. The latter will accommodate a structure similar  to an isoelectronic  elemental structure. The resulting structure  contains  isolated cations and oligomeric or polymeric anions (Zintl anions) which can form 0-D, 1-D, 2-D or 3-D arrangements.  The cations are understood being a template whose main role is the electrostatic stabilization whilst the  Zintl anions determine the electronic structures\cite{Wengert,Nesper6,Nesper7,Nesper9}. ZP have relatively high melting points, are brittle  and often diamagnetic. In the last decades, a great variety of these compounds have been synthesized under inert atmosphere; e.g.  BaSi${}_{2}$, Ba${}_{3}$Si${}_{4}$, BaSi, Ba${}_{5}$Si${}_{3}$, Ba${}_{2}$Si \cite{Currao,Zurcher}. Their high reactivity with air and water can be understood from the frequently  odd  oxidation states of their constituents, the highly negative charges on the Zintl anions and the resulting strongly reducing properties. On this same line, one could also suspect unusual electronic properties; a reason for which their study is worthwhile\cite{Reinhard91}. Some Zintl compounds, which contain ecliptically stacked planar silicon chains, have been found  to  exhibit metallic properties\cite{Nesper8} with a very weak temperature independent paramagnetism. Yet, theoretical calculations\cite{Wengert} have proved that their transport properties are  determined by the Zintl anion substructure.

Alkaline earth monosilicides ({\AE}Si), with {\AE}= Ca, Sr and Ba, are  examples of ZP which crystallize with the CrB structure; space group  $Cmcm$ N${}^{\circ}$ 63 \cite{Rieger,Eisenmann,Merlo}.  The silicon substructure fulfills the geometrical conditions for a six valence electron species, i.e. sulfur. Formally, Si$^{2-}$ anions form eclipsed zigzag chains, ${}^{1}_{\infty}$[Si$^{2-}$], and {\AE}$^{2+}$ cations are isolated; meaning that silicon derived bands may be considered to be full.  

Studies of electronic structure have been carried out on several alkaline-earth silicides\cite{Watanabe,Brutti} but there is, to our knowledge, no complete investigation of the properties of the phases with the 1/1 composition. The planarity of the Zintl anion, with silicon in a  $2-$ valence state, is intriguing and suggests that there might be a significant direct electronic coupling between the chains.

As with other compounds with ecliptically stacked silicon polyanions\cite{Nesper8}, we have found {\AE}Si  to have high electric resistivity  with  a positive and constant temperature gradient, indicating the absence of a band gap.   \ref{Resistivity}  exhibits data of  four point van der Pauw measurements. For the sake of comparison, we have also included  the resistivities of  the individual elements in question. 

\begin{table}[h]
\tabcolsep=\tabcolsep
\begin{center}
\caption[Resistivity data at room temperature.]{Resistivity data at room temperature.}\label{Resistivity}
\begin{tabularx}{0.6\textwidth}{l r @{.} l r @{.} l  l r @{.} l }
\toprule
 {\AE}Si & \multicolumn{2}{c}{Resistivity }  & \multicolumn{2}{c}{$\frac{1}{\rho}\frac{\Delta\rho}{\Delta T}$}  & Element & \multicolumn{2}{c}{Resistivity }  \\
 & \multicolumn{2}{c}{ ($10^{-4}_{}\Omega$ $\cdot$ cm)}   &  \multicolumn{2}{c}{(K$^{-1}$)}  & & \multicolumn{2}{c}{ ($10^{-4}_{}\Omega$ $\cdot$ cm)}  \\\midrule
 CaSi   &    0 & 618    &  0&0001                    &Ca   &    0 & 034\\
 SrSi   &  127 & 6      &  0&004                     &Sr   &    0 & 13 \\
 BaSi   &   68 & 2      &  0&004                     &Ba   &    0 & 35 \\
        &\multicolumn{2}{c}{} &  \multicolumn{2}{c}{}&Si   & 1000 & \\
 \bottomrule 
\end{tabularx}
\end{center}
\end{table}

In this paper we show that  single crystals of {\AE}Si have unusual magnetism. This is interesting, not only because it contradicts the expected behavior of classical Zintl compounds, but also because a collective magnetic response is generally related to the exchange interaction involving localized electrons in incomplete $d$ or $f$ shells. Here, we present the main results of our magnetic measurements as well as theoretical results, based on first-principles calculations,  which did help us to  propose an explanation for the observed effects. A detailed study of the electronic structures of {\AE}Si will be presented elsewhere in the future\cite{EdRein}.

 \section{Methods}
 \subsection{Preparation of the samples}
 Calcium 99.5 wt. \%, strontium 99.9 wt. \% and barium 99.9 wt. \% (Alfa Aesar GmbH) metals were redistilled twice under inert conditions. Then, dendritic pieces of the metals were mixed with silicon powder 99.995\% (Alfa Aesar GmbH) and enclosed in a tantalum tube under purified argon atmosphere. Polycrystalline samples were prepared by annealing at 1000 K for 2 weeks and subsequently quenched mixtures of starting materials (starting mass 2-2.5 g) were pressed into pellets and melted on a water-cooled copper holder using light-arc furnace. The temperature of the arc was slowly decreased to achieve slow melt crystallization. Resulting samples in the form of brittle bulk of prism-like dark-gray crystals were analyzed by means of X-ray diffraction (Bruker AXS D8, CuK$\alpha$1 radiation) using a protective argon-filled sample holder.  For magnetic measurements, single crystals were separated  and placed in quartz tubes under inert atmosphere. Later, the composition of  these single crystals was analyzed by laser ablation inductively coupled plasma mass spectroscopy (LA-ICP-MS). This technique allows a quantitative determination of impurities in  single crystals at different depth and at concentration levels down to the ppm range\cite{Guenter}. 
 \subsection{Magnetic measurements} The measurements were done with a Superconducting Quantum Interference Device (SQUID) from Quantum Design\cite{McElfresh}. The SQUID has a superconducting magnet which can operate  an applied field up to 5 T and measures moments ($m$) as low as 10$^{-8}$ emu (10$^{-11}$ Am$^2$). It operates over a temperature range from 1.9 K to 400 K. We used a scan\footnote{The sample, being much smaller than the length of the second-order detection coil, is moved along the axis of the latter, which detects the change in the magnetic induction produced by the sample. The length of this scan has to be larger than the sample size, but short enough to avoid inhomogeneities of the applied field.} of 3 cm. The magnetic moment reported for each measurement is the average of three scans and  every point of each scan was measured ten times. Over all, the standard deviations of the  moments  are two or more orders of magnitude smaller than their average  values. Measurements are considered reliable when  $m>10^{-8}$ emu for temperature scans but  two orders of magnitude higher ($m>10^{-6}$ emu) over field loops.

\subsection{First-principles calculations}
Electronic structure calculations presented in this paper were performed using the CASTEP\cite{Castep} code from Materials Studio and the code based on linear-muffin-tin-orbitals (LMTO) from Andersen's group\cite{Andersen,AndersenTB,AndersenTB1,AndersenTB2,AndersenTB3,AndersenX1}. 

\begin{figure}[h!]
\vspace{0.005\textheight}
\includegraphics[width=0.6\textwidth]{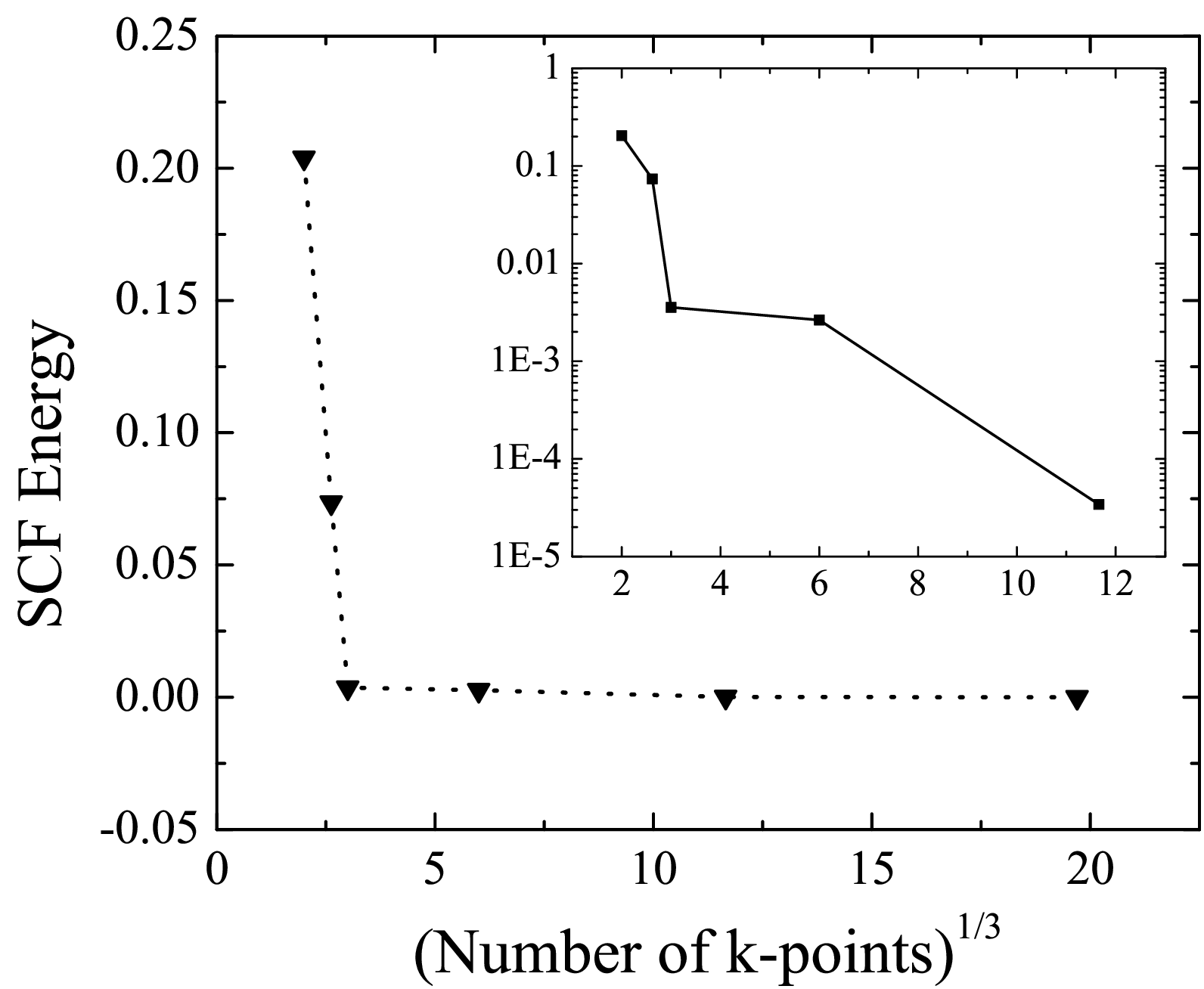}
\caption{Convergency test. Self-consistent field energy {\em vs } qubic root of the number of $k$-points, for BaSi. In the main frame: difference with respect to the energy using $20^3$ points; in-set: the same in log-scale.}\label{energyconverg}
\end{figure}

Within CASTEP, we took ultrasoft  pseudo-potentials and we  used the generalized gradient approximation (GGA) with the Perdew-Burke-Erzenhof functional (PBE)\cite{Perdew2}. Pseudo-atomic calculations were done employing the reference configuration Si $3s^2_{}3p^2_{}$ and {\AE} $ns^2_{}np^6_{}(n+1)s^2_{}$ with $n=$ 3, 4, and 5 for Ca, Sr and Ba, respectively. Thus, every selfconsistent calculation was done with 56 electrons. The CASTEP code applies a plane wave basis set. The plane wave cut-off was set to 310 $e$V and the $k$-point sampling was done on a $6\times 6\times 6$ mesh following the Monkhorst-Pack procedure\cite{MP}. We tested the convergence of the total energy as a function of the number of $k$-points (shown for BaSi in \ref{energyconverg}). We used  the default value of 0.1 $e$V for the Gaussian smearing of the electronic states;  systems were treated as metallic with variable occupancy and the number of empty bands was increased to 30\%. In order to test the robustness of our results, we tried different self-consistency algorithms within the CASTEP package. The tolerance for selfconsistency was set to $0.5\times 10^{-6}_{}$ $e$V  for the total energy per atom, $10^{-7}_{}$ $e$V for the eigenvalues and $10^{-8}_{}$ $e$V for the Fermi level. Lattice constants and atomic positions of {\AE}Si are well known but still we checked for possible geometry optimization. The structures were optimized with a tolerance of $10^{-2}$ $e$V/{\AA} for the forces, $5\times 10^{-6}$ $e$V for the change in energy per ion and $5\times 10^{-4}$ {\AA} for the displacements. Experimental and optimized atomic positions differ in less than $10^{-3}_{}${\AA} and the difference in total energies are of the order of 10$^{-3}e$V. Therefore, the experimental structures can be considered optimized.

LMTO calculations were performed using the atomic sphere approximation (ASA), version LMASA-46, and in order to minimize the overlap, empty spheres were introduced with radius between 1.0 and 4.5 a.u.. The selfconsistency convergence tolerance was set to $10^{-5}$ Ry for the total energy and $10^{-5}$ $e$ for the RMS of the change of atomic charges. For silicon atoms, orbitals ($nlm$) with $n<3$ were treated as core states and, for {\AE} atoms, atomic orbitals with energy below the valence  $s$ orbital  were considered core states. $l=2$ was the highest partial wave considered. For the selfconsistency algorithm, Broyden mixing was used. The $k$-space was sampled on a grid of $16\times 16\times 16$ points, and the  integration was done using the tetrahedral method\cite{AndersenY1,AndersenY2}.  Systems were treated as metallic and the Fermi level was determined with an accuracy of $10^{-6}$ Ry. A very useful feature of the LMTO method lacking for other methods, is that the charge distribution into muffin tin orbitals is  physically meaningful. Most methods use a ``rigid'' basis set (atomic orbitals) and solve the structure by finding the best coefficients to  represent the density by a linear combination of atomic orbitals (LCAO). However,  atomic orbitals in a solid have much less physical meaning than in an atom since electrons are shared by multiple centers. Therefore, the intent to distribute the density onto LCAO often results in misleading occupations. In the LMTO formalism, the muffin-tin orbitals are constructed in the self consistency algorithm. The tail cancellation  and the continuity conditions imposed on the muffin-tin orbitals produce local states which are intrinsic to the solid  rather than of atomic origin. Therefore, the occupation of  muffin-tin orbitals gives a better representation of the electron distribution than  projections onto other LCAOs.

\section{Results and discussion}

\begin{figure}[h!]
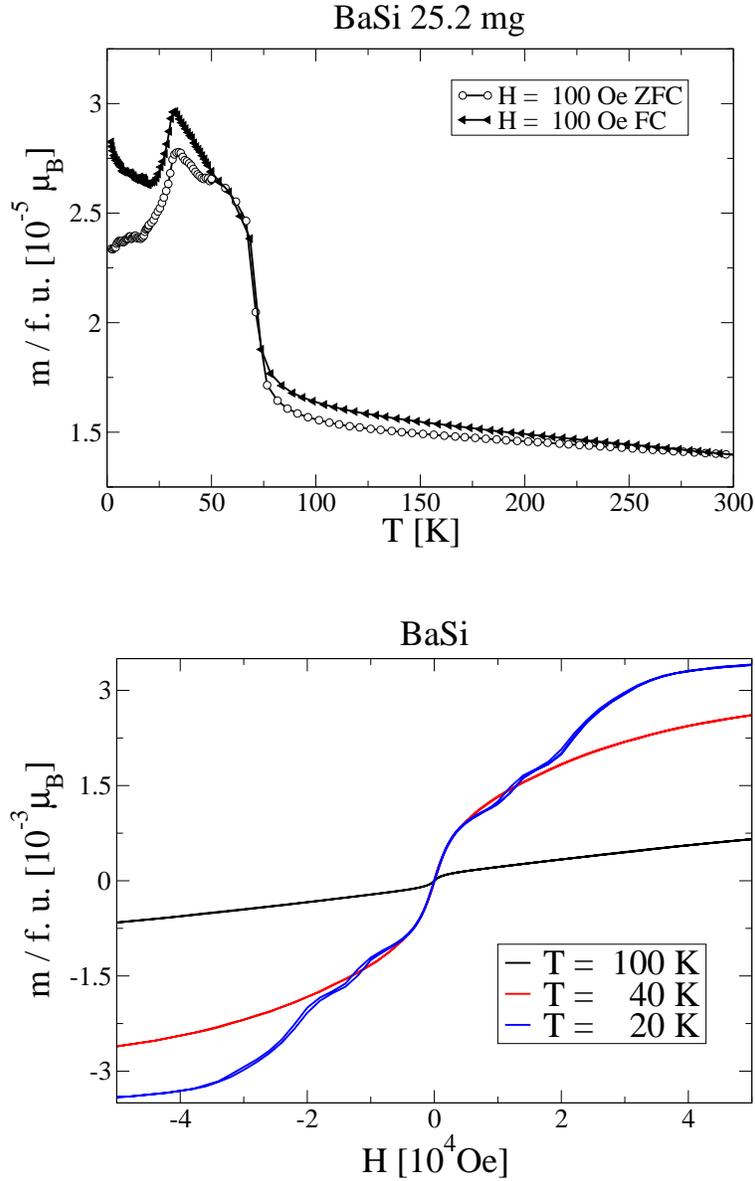

\vspace{0.01\textheight}
\includegraphics[width=0.6\textwidth]{BaSiH100MvsT.eps}\\
\vspace{0.04\textheight}
\includegraphics[width=0.6\textwidth]{BaSiMvsH.eps}
\caption{Magnetic moment {\em vs} Temperature (up) and {\em vs} applied field (down)}\label{MvsHBaSi}
\end{figure}
The upper panel in  \ref{MvsHBaSi} shows the magnetic moment per formula unit {\em vs} temperature for a single crystal of BaSi, under a field of 100 Oe. ZFC (zero field cooled) designates measurements done while warming a sample that was previously cooled without magnetic field present. FC represents the measurements during cooling down in the presence of a field. The high temperature part of this graph shows a nearly temperature-independent paramagnetic response. A ferromagnetic-like  transition is visible around 70 K with a sudden increase of the magnetic moment and splitting of  ZFC and FC curves. The cusp at 35 K is typical of antiferromagnetic ordering. The coexistence (or competition) of these two  effects suggests the presence of different exchange couplings. It resembles  spin-glass behavior caused by frustrated interactions. The lower panel of \ref{MvsHBaSi} shows the change from linear to non-linear field dependency of the magnetic moment per formula unit when it is measured at 100 K, 40 K and 20 K. Although all results obtained from different samples are qualitatively similar, the observed moments and transitions are small and the actual values depend on the synthesis. This will be commented on later.

\begin{figure}[h!]
\vspace{0.005\textheight}
\includegraphics[width=0.4\textwidth]{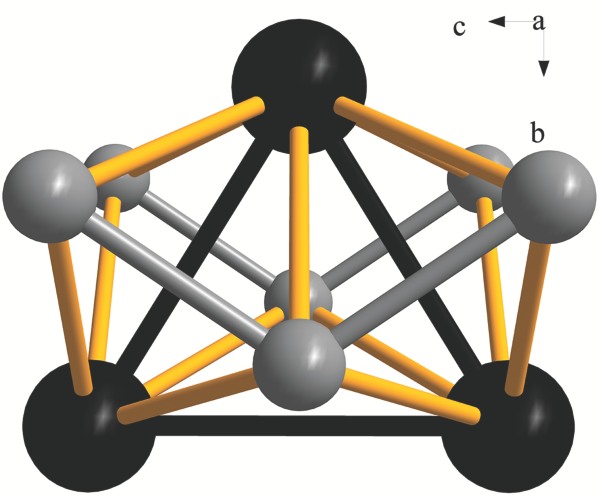}
\caption{Triangular coordination of metal (black) atoms. Base, 4.14 {\AA};  side, 3.93 {\AA}. Inter-triangle distance (along $a$), 5.04 {\AA}}\label{coordination}
\end{figure}

The direct exchange interaction between electrons in valence states centered at different atoms is negative\cite{Zener53} while exchange mediated by ligands can have negative or positive sign depending on several factors. \ref{coordination}  shows the relevant coordination of the atoms in the {\AE}Si structure.  Provided that there were electrons in the $d$ states of metal atoms, the observed transitions could be related to frustrated  exchange interactions within the triangular prismatic arrangement. The stepwise (metamagnetic) $M$ {\em vs} $H$ dependence would correspond to the field induced spin-flip happening when the Zeemann energy becomes similar to one of the antiferromagnetic couplings.  However, the valence states of the metal atoms are expected to be nearly empty according to the Zintl concept. Moreover, $d$ states of  {\AE} metals are extended and, if occupied, they should overlap and mix with  states of neighboring atoms. {\AE}Si should not contain local moments and  exchange interactions are expected to be very weak. 

\begin{table}[h]
\begin{center}
\caption{Room temperature susceptibility of MSi [M] at  a field of 1 T,  and saturation moments  under ${}^{\rm (un)}_{}$ and above ${}^{\rm (ab)}_{}$ 75 K.}\label{resumechi}
\begin{tabularx}{0.56\textwidth}{rr@{.}lr@{.}lr@{.}lr@{.}l}
\toprule
 MSi &  \multicolumn{4}{c}{$10^{4}_{}$ $\chi^{}_0$ }  & \multicolumn{2}{c}{$M^{\rm (un)}_{\rm S}$ per f.u.} & \multicolumn{2}{c}{$M^{\rm (ab)}_{\rm S}$ per f.u.}\\
    &  \multicolumn{4}{c}{( cm${}^{3}_{}$ $\cdot$ mol${}^{-1}_{}$)}  & \multicolumn{2}{c}{($10^{-5}_{}\mu^{}_{\rm B}$  )} & \multicolumn{2}{c}{($10^{-5}_{}\mu^{}_{\rm B}$  )}\\
\midrule
 CaSi   &  0&1   &  [0&4]                        &  2&69 &  2&15 \\
 SrSi   & -0&6   &  [0&9]                     & 16&11  &  $<$0&08 \\
 BaSi   & -0&25  &  [0&2]                      & 196&90 & 15&22 \\
 ScSi   &  0&2   &  [2&9]                   & 1&60\textsuperscript{\emph{?}} & 1&61\textsuperscript{\emph{?}}  \\
 YSi    &  0&5   &  [1&9]                      & 2&86 & 2&41\\
 LaSi   &  0&8   &  [1&0]            & 1&87\textsuperscript{\emph{?}}  & 0&17\textsuperscript{\emph{?}}\\
\bottomrule
\end{tabularx}\\
\textsuperscript{\emph{?}}No reliable data.
\end{center}
\end{table}

 \ref{resumechi} shows the high temperature susceptibility data  of the mono-silicides. Their low $\chi^{}_0$ values and high resistivities (from \ref{Resistivity})  indicate that they are poor metals.  \ref{resumechi} also shows the saturation moment  $M^{}_{\rm S}$ below and above 75 K.  Amongst the analyzed samples, $M^{\rm ab}_{\rm S}$ is  remarkably  different to $M^{\rm un}_{\rm S}$  in SrSi and BaSi. For ESi (E = Sc, Y), which also crystallize with the CrB structure type and have one extra valence electron with respect to {\AE}Si, $M^{}_{\rm S}$ are very small and rather constant over the temperature explored. This indicates that the reduction of  size of the $d$-states, with respect to those of {\AE} metals, together with the addition of one electron, do not enhance the observed magnetism.

 \begin{figure}[h]
 \vspace{0.03\textheight}
\includegraphics[width=0.6\textwidth]{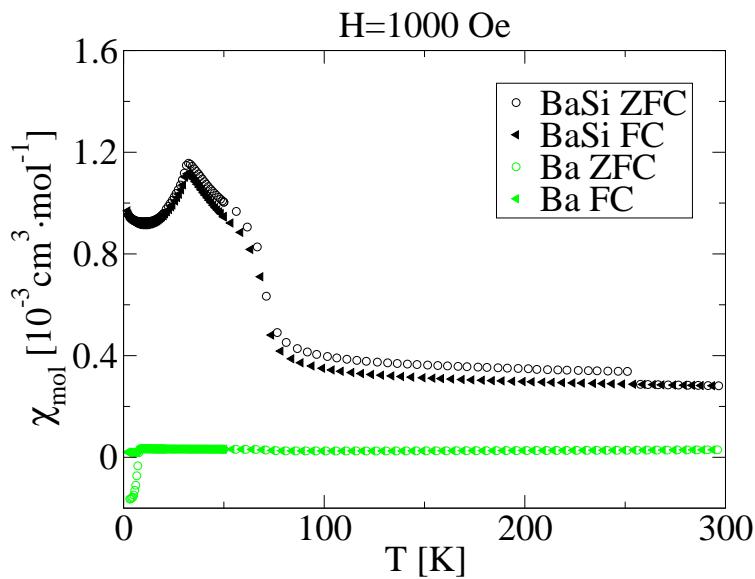}
\caption{Comparison of $\chi(T)$ of BaSi and Ba used as starting material}
\label{MetalandMetalsilicides}
\end{figure}

Considering these evidence, a first check was the analysis of impurity effects to ensure that the magnetism is intrinsic to  crystalline compounds or due to doping by magnetic centers.  Crystallization is known to be a purification mechanism, very effective in Zintl phases. In single crystals of Zintl compounds, impurities will be preferably located as single atoms replacing   {\AE} or Si atoms in their crystallographic positions, or as surface nano crystals.  In order to judge the role of impurities, the  composition of the single crystals used in the magnetic measurements was  analyzed by LA-ICP-MS. The impurities (excluding substitution by other {\AE}) were in all cases under $10^{-5}_{}$ atoms per formula unit and evenly distributed at different depths  with respect to the crystal surfaces\cite{Guenter}.  The corresponding impurity densities are too low to produce collective magnetic effects observed in the SrSi and BaSi samples.   In addition, we tested the magnetic response of the starting materials.  \ref{MetalandMetalsilicides} compares the response of a single crystal of BaSi and a piece of Ba used as starting material, confirming  that the magnetism is intrinsic to the silicide. Similar tests on samples of  silicon used as starting material show  perfect diamagnetic response with a temperature independent susceptibility of approximately $-1.1\times 10^{-5}_{}$ cm${}^{3}_{}$ mol ${}^{-1}_{}$.

With these experimental evidence and the fact that these systems are poor metals which do not contain localized moments, our best explanation of the observed magnetism is the polarization of the low density of charge carriers. The Coulomb interaction always favors the formation of magnetic moments but it is generally hindered by the dominant kinetic energy, which is minimized when bands with up- and down-spin are equally filled. In a common metal, the free-electron density is too high to be polarized\footnote{The mechanism that produces magnetism in elemental metals (Fe, Ni and Co) is known from Zener works\cite{Zener51a,Zener51b} as a double exchange where the conduction $s$ electrons interact via local exchange with the incomplete $d$-shell. The scattering of the conduction electrons is minimized when the local moments are aligned. In other words, localized moments are coupled through the conduction band.}. For low electron-densities, the increase in kinetic energy caused by the partial polarization of the electrons can be overcome by the gain in exchange between itinerant electrons; in this way, the paramagnetic state can be unstable with respect to magnetic ordering of mainly itinerant type\cite{Ceperley}. To make this hypothesis sustainable, we performed first-principles electronic structure calculations. 
 
\begin{figure*}[ht]
\includegraphics[width=\textwidth]{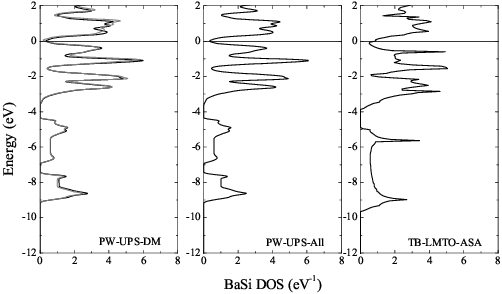}
\caption{Total DOS. Plane waves basis using density mixing algorithm (left), plane waves basis using  standard DFT algorithm (central), LMTO method (right).}\label{4DOSBaSi}
\end{figure*}

\begin{figure*}[ht]
\includegraphics[width=\textwidth]{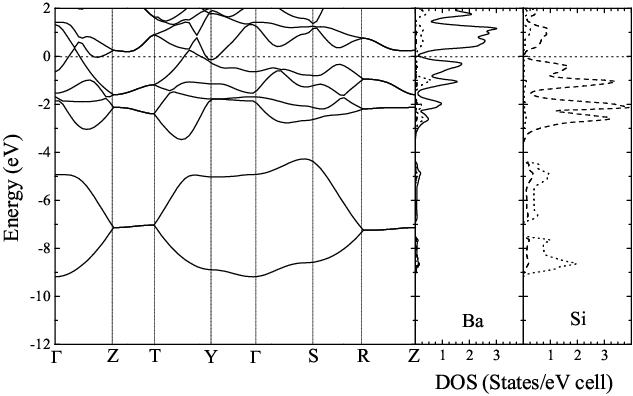}
\caption{BaSi bands and DOS. Dotted, dashed and solid:  $s$, $p$ and $d$ states, respectively}\label{BNDBaSi}
\end{figure*}
 
 At the DFT level, one can not be very optimistic about  results concerning magnetic states with polarizations as small as observed in the experiments, involving weak and competing exchange interactions. DFT is known to underestimate exchange effects, mostly those originating from non-local interactions. However, the electronic structures of these  weakly magnetic materials are expected to differ only slightly from the paramagnetic solutions. Therefore, one should be able to see within the DFT level whether or not a poor metallic phase is occurring and  where the corresponding electrons arise from.

The results of our calculations are stable and consistent as can be seen in  \ref{4DOSBaSi} where the total DOS of BaSi resulting from several methods are compared. The left panel contains the results form CASTEP calculations using density mixing in the  selfconsistent algorithm, the central panel shows the CASTEP results using the standard DFT algorithm, i.e. optimizing the total energy with respect to all coefficients of the wave function, and the right panel depicts the DOS obtained with the framework of the LMTO scheme. In all cases BaSi  is predicted to be a poor metal.

\begin{figure*}[ht]
\includegraphics[width=0.6\textwidth]{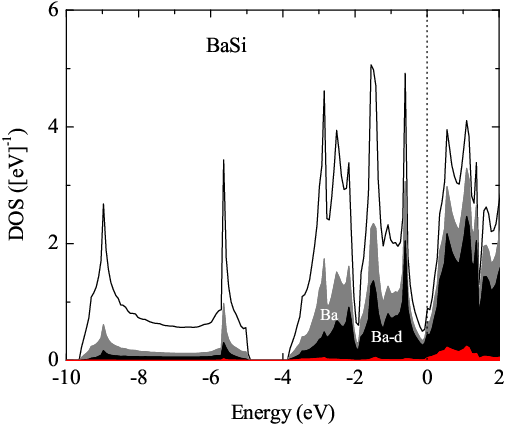}
\caption{DOS from LMTO. White area (total DOS), grey area (Ba DOS), black area (Ba-$d$ DOS), red area (empty spheres DOS).}\label{LMTODOS}
\end{figure*}

Taking a closer look,  \ref{BNDBaSi} shows the band structure and partial DOS of BaSi obtained with the CASTEP package.  The small overlap of conduction and valence bands is a consequence of the broadening of the $\pi$-system band arising from eclipsed stacking of the silicon chains.  According to the projection of the plane-waves basis onto LCAO, there is a significant contribution of metal $d$ states below $E^{}_{\rm F}$. With the projector we used, the identity difference and the spilling parameter are small -- both around $10^{-3}_{}$ -- which prove that the projector can be considered complete  and the occupation of excited states of metal atoms is not an artifact of the calculation as a consequence of the under representation of the electronic structure.  

  The population of $d$ states has been confirmed by  LMTO calculations (\ref{LMTODOS}). This participation of $d$ orbitals and the low DOS($E^{}_{\rm F}$) make  the occurrence of magnetic states plausible. Together with the $\pi$ states, $d$-orbitals form a conduction network with mainly 2-dimensional character. The extended $\pi$ states serve as  long-range links between the $d$ orbitals and the low electron density makes the system polarizable and also very sensitive to defects and impurities, which can explain the sample-dependence of the experimental results.  The participation of $d$ orbitals  makes  the triangular prismatic arrangement of the metal atoms reasonable by means of metal-metal interactions while the planarity of the silicon substructure becomes supported by the deficit of electrons in the $\pi$-system due to this transfer towards metal states thus leading to incompletely occupied $\pi$-bands and a net $\pi$ bonding contribution. 
   \begin{figure}
\includegraphics[width=0.6\textwidth]{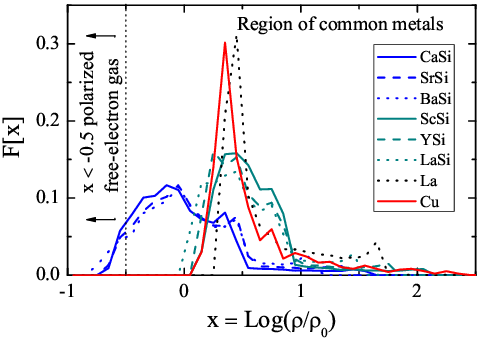}
\caption{Histogram of the valence-electron density comparing silicides with some metals}\label{distrib}
\end{figure}

\ref{distrib} shows the log-distribution of effective valence electron density for {\AE}Si, ESi, Cu and La; last two as examples of typical metal densities. The distribution in  {\AE}Si is broad in comparison to common metals\footnote{For a structureless system, free electron gas, $F$ is a delta-function.} and enters the region  where the free electron gas has been predicted to be magnetically unstable, according to Montecarlo calculations \cite{Ceperley}. Earlier in this manuscript, we mentioned that the observed moments vary from sample to sample and that a low carrier concentration could be an explanation for this. We also argued that the magnetism can not be just caused by impurities because they were only found at very low concentrations. Nevertheless, whether the magnetism will persist or not in an ideal crystal (which has neither impurities nor defects at all) is a question that we can not yet answer. Defects or impurities could play a role destabilizing the paramagnetic state for low density of carriers,  while these effects might not take place in the starting materials with similar amount of impurities. We are convinced that the observed magnetism is not due to an impurity band since these silicides are found to be intrinsically metallic. Further more we argue that  due to the measured impurity densities, the effective radius of the impurity orbital would have to be very large ($a^{*}_{\rm B}\approx 60$ {\AA}) to reach the percolation threshold.

 We performed spin polarized calculations with several  initial settings for the spins. We took $1+$ as initial charge for {\AE} atoms and we let the electronic structure to relax towards selfconsistency.   The calculated net moment ($\sum_i S^{}_i$) per unit cell  depends on the initial settings (whether initial spins were set all parallel or the orientation was alternated in different crystallographic directions), taking values between $10^{-7}_{}$ $\mu^{}_{\rm B}$ and  $10^{-4}_{}$ $\mu^{}_{\rm B}$ per formula unit and, $\sum_i |S^{}_i|\sim 10^{-3}_{}$ $\mu^{}_{\rm B}$ in all cases.  Differences in total energy per cell between any pair of settings are smaller that $0.014$ $e$V. These energy differences are very small but they are still larger than the errors and larger than the energy differences between experimental and optimized structures. These different solutions point to the existence of several nearly-degenerate minima.  The degeneracy of the ground state  is  consistent with the  frustration of the spins and the glassy behavior, but we think that further studies including non-local exchange interactions and/or with a higher resolution of the DOS at the Fermi level are necessary for a complete understanding of such systems.  These computationally demanding tasks  can be hopefully carried out in the near future.
%

\section{Conclusions}
In this paper we have shown that alkaline earth mono-silicides do not behave as classical Zintl phases, in contrast with their ideal composition. They are poor metals due to the transfer of electrons from the broad $\pi$- system of the eclipsed silicon chains toward  $d$-states of {\AE} atoms. This is consistent with the planarity of the chains, the triangular prismatic coordination of metal atoms and in conjunction, they seem to give room for weak magnetism to occur. Most likely this magnetism is of itinerant type, yet with frustrated exchange interactions. In the near future we will study the role of the dimensionality  and the shape of the density of states. It would be also interesting to check how theoretical predictions change when non-local exchange interactions are taken into account.  Systematic experiments aiming to quantify the role of impurities are also being considered.

\begin{acknowledgement}

The authors thank Prof D. G\"unter's  group at ETH-Zurich  for the elemental analysis of our samples. This project has been supported by the Swiss National Science Foundation under project no. 2-77937-10.
%
%
%
\end{acknowledgement}

\end{document}